\documentclass[noshowpacs,amsmath,amssymb,twocolumn]{revtex4}

\usepackage{graphicx}
\usepackage{amssymb}
\usepackage{amsmath}
\usepackage{wasysym}
\usepackage{gensymb}

\begin{document}
\title{Minimalist design of a robust real-time quantum random number generator}

\author{K. S. Kravtsov}
\email{kravtsov@kapella.gpi.ru}
\author{I. V. Radchenko}
\affiliation{Faculty of Physics, Moscow State University, Moscow, Russia}
\affiliation{A.M. Prokhorov General Physics Institute RAS, Moscow, Russia}
\author{S. P. Kulik}
\affiliation{Faculty of Physics, Moscow State University, Moscow, Russia}
\author{S. N. Molotkov}
\affiliation{Academy of Cryptography of Russian Federation, Moscow, Russia}
\affiliation{Institute of Solid State Physics, Chernogolovka, Moscow Rgn., Russia}
\affiliation{Faculty of Computational Mathematics and Cybernetics, Moscow State University, Moscow, Russia}

\begin{abstract}
We present a simple and robust construction of a real-time quantum random number generator (QRNG). Our minimalist
approach ensures stable operation of the device as well as its simple and straightforward hardware implementation as a
stand-alone module. As a source of randomness the device uses measurements of time intervals between clicks of a
single-photon detector. The obtained raw sequence is then filtered and processed by a deterministic randomness
extractor, which is realized as a look-up table. This enables high speed on-the-fly processing without the need of
extensive computations. The overall performance of the device is around 1 random bit per detector click, resulting in
1.2 Mbit/s generation rate in our implementation.
\end{abstract}

\maketitle

Random number generation attracted a lot of attention since quantum technologies became available to serve for this
purpose. Quantum random number generation gives access to the universal randomness of quantum mechanics, which is
unavailable, at least theoretically, in classical physics. Over the last 20 years there was a number of different QRNG approaches
and suggestions about its practical implementation.  In the present work we looked for a
highly practical and easy to implement QRNG, which will ensure stable operation due to its overall simplicity. To keep
it simple we intentionally considered only designs with one single photon detector. Along with the physical part, not less
important was the choice of a randomness extraction algorithm, which, by itself, should guarantee generation of
independent and unbiased bits at the output under any conditions.

There are many different approaches to quantum randomness generation described in the
literature~\cite{V70,JAW00,SGG00,FSS07,DYS08,WJA09,KAR09,GWS10,WLB11,SMZ14,NZZ14}. Still, it is hard to
clearly distinguish between what we call ``quantum'' and what in some sense belongs to statistical physics. We associate
the former with the use of simple ``quantum measurements'' and single photon (or other particles) detectors (SPDs), and
believe that most other methods belong to the latter case, where it is harder to explain why the source is truly
unpredictable and thus whether it can in principle generate true randomness.  Among the SPD-based sources it is highly
beneficial to use a single SPD in the whole setup. Besides the obvious reliability advantage, it also eliminates the
need of any hardware calibration as SPDs typically have a large variation of performance parameters even within a
particular type. Using more than one SPD also does not give an advantage in the generation rate: typically the result is the
same as the corresponding multiple of single SPD devices. 

The output of an SPD contains only timing information (it could be also a number of photons, but conventional SPDs do
not resolve it) i.e. moments when a particular photon caused an avalanche in the detector. This is the only data available for
random numbers extraction.  Realization of a quantum measurement may require engineered light pulses,
for example in a ``which way'' experiment~\cite{SGG00}, or in a ``multiple choice'' one~\cite{LCL07}, but from the
standpoint of performance
those settings are suboptimal: instead of using any time moment measurable by the clock,
we intentionally limit ourselves to only a few or several of them. Thus, in this sense optimal design should use
a continuously running light source.

The simplest form of this method is detection of the arrival times of photons from a cw coherent
source~\cite{V70,DYS08,WJA09,WLB11,NZZ14}. This gives
purely Poissonian statistics, which is simple to work with. A modification with temporally shaped cw source is also
possible~\cite{WK10} and gives slightly better performance at the expense of an increased system complexity and reduced
reliability, which is unsuitable for our concept.

Collecting timing information is commonly done in two different forms: timestamping arriving photons by the value of a
free-running clock counter as in~\cite{V70,DYS08,FWN10,NZZ14} or counting the number of clock cycles between adjacent
detection events~\cite{WJA09,WLB11}.
The two are completely identical by the amount of obtained information (each form of data can be easily converted to the
other one), but the second form consists of independent integers, while the first one unavoidably contains internal
correlations. The correlations come from the slowly changing most significant bits of the time counter. A direct
calculation~\cite{V70} shows that in order to make correlations negligible, one has to ensure that bit values change some 15 times faster
than the average counting rate, which poses a significant limitation on the amount of random data obtained. Thus,
storing time intervals between detector clicks is advantageous as, by the definition of the Poisson process, all such
values are perfectly independent.
They are distributed with the exponential distribution, favoring small numbers and exponentially decaying
towards the large ones.

Another not less important aspect of this approach is light generation. In principle, it would be the best to use
a single (transverse, temporal, and polarization) mode cw coherent light source. In practice, however, it is not
possible. We use instead an alternative approach, based on mode averaging. Even if the statistic of a single mode is not
necessarily Poissonian, the more modes we detect the closer the output will resemble Poissonian process due to the
Palm–Khintchine theorem~\cite{HS04}. Thus, a substantially multimode detection allows to arrive at a Poissonian process regardless of
the photon statistic in each of the modes.

After a raw stream of random data is acquired one needs to process it in order to obtain random and independent bits.
This processing, or extraction, can be done in many different ways with two global approaches: using deterministic
algorithms and with seeded extractors. The second approach is much more powerful as it is capable of extracting
randomness from any type of a partially random sequence, however, in this work we stick to the first one. Seeded extractors
need a perfectly random seed to operate. Until recently such algorithms required more random bits than they
produced, e.g. using Toeplitz matrices~\cite{NZZ14}. Later, a new Trevisan extractors family~\cite{T01} was introduced, which
allows for generating more randomness than was consumed as the seed. It also allows re-using the seed for
future conversions~\cite{MXX13}. 

The main difficulty with seeded extractors is that they are constructed for a particular rate of min-entropy provided by
the source. After it is defined the extractor gives a certain rate of compression, which does not depend on anything,
including the actual quality of the raw sequence. This creates a possible flaw in the system: if by any reason the raw
sequence degrades, the output of the seeded extractor will cease to be truly random.
Such algorithms intrinsically lack any adaptivity and may easily fail with a temporal
drift of experimental parameters affecting the amount of generated entropy.

Other problems include estimation of min-entropy of the source and real-time realization of such, especially
Trevisan, extractor. Strictly speaking, it is impossible to measure min-entropy of any given source in finite time.
Knowing the internal structure of the source helps but does not make the problem trivial. At the same time this
parameter remains the only parameter that defines whether the QRNG output is random or not. It may become even more
critical with re-using the extractor output as the seed.
Real-time processing is rather a technical but nevertheless one more important issue. Recent results show that realization of
Trevisan construction with modern computers yields some 17~kbits/s output bitrate~\cite{MPS12}. Apparently, this is orders of
magnitude slower than the entropy generation rate by a typical SPD. Such computational-hungry algorithm is hard to run
when the system performance is not ignored.

\begin{figure}
\centering
\includegraphics[width=\columnwidth]{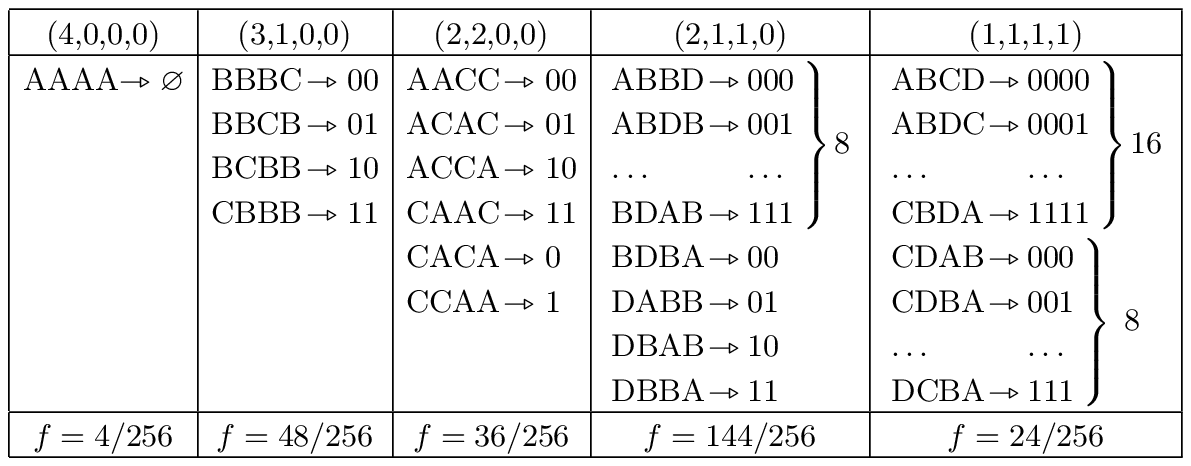}
\caption{Extraction of random bits for $M=N=4$. The first line shows a word pattern, i.e. a sorted list of repetition
numbers of each letter. The corresponding extraction process is given below on the example of one permutation family.
The value $f$ shows the share of a particular pattern in all $M^N =256$ possible input words; it coincides
with the actual frequency of appearance if all letters are equiprobable.
}
\label{fig_conversion}
\end{figure}

Based on the given discussion, we think that deterministic algorithms, if applicable, are better suited for our goals.
Here we should notice that a popular approach~\cite{WJA09,WK10,GWS10} of using standard hash functions such as SHA-256 is, strictly speaking,
invalid~\cite{WLB11} as it does not extract randomness, but rather only compresses the raw data in some peculiar fashion. It does not
guarantee independence of output bits and possesses similar weaknesses as seeded extractors. Talking about deterministic
extractors, one needs to know properties of the input data. As was shown earlier we consider a stream of
positive and independent integers with the exponential probability distribution. The goal of the extractor is to convert
this stream into unbiased and independent bits.

The most trivial but inefficient method is to convert the stream into independent but biased bits by taking the integers
modulo 2 and then eliminate the bias using the von Neumann algorithm~\cite{N51}. We use a generalization of this method to a finite
alphabet of  $M>2$ letters --- the extended Elias algorithm~\cite{E72}, which is asymptotically efficient, converging to the entropy of
the source (see also~\cite{P92}). 
Importantly, here any deterministic conversion of the obtained positive integers into the finite alphabet may be used, as it does not
affect the quality of the extractor output. It is desirable, however, to approach the uniform probability distribution within
the alphabet to decrease the entropy loss.

\begin{figure}
\centering
\includegraphics[width=\columnwidth]{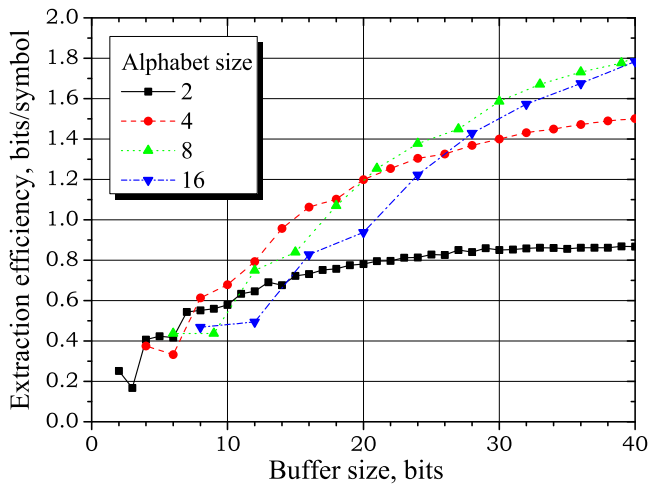}
\caption{Extraction efficiency of the implemented algorithm vs. the size of the buffer for different alphabet sizes. The
actual generator uses a 20-bit buffer and a 4-symbol alphabet, yielding 1.2~bits per raw symbol.
}
\label{fig_extraction}
\end{figure}

Realization of the Elias extraction algorithm is explicitly given in~\cite{JJS00}. Briefly, the input sequence is cut into words
of $N$ letters each. A probability of getting a particular word is exactly the same as for all the words, which are
permutations of letters of the original one. The total number of such permutations gives the number of equiprobable
states, one of which is obtained as the original word. All such states can be numbered, and the number of the one
obtained can be converted into random bits using its binary representation.

An example of extraction for $M=N=4$ is shown in Fig.~\ref{fig_conversion}.
The strings AAAA, \dots, DDDD are mapped onto the empty set because they all have different probabilities of appearance
unless the probability of obtaining the individual letters A, B, C, and D are exactly equal. The patterns ABBB, BABB,
BBAB, and BBBA, and the like, generate two output bits because if the letter-generating events are independent, each
combination of letters has 4 equiprobable permutations irrespective of the probability of obtaining the individual
letters A, \dots, D. The largest number of permutations is obtained for the word ABCD, giving 24 variants; each results in 3
or 4 output bits.

Following our minimalist doctrine we wanted to simplify required processing as much as possible. Taking into account
that the expected throughput of the extractor should be in Mbits/s range, we sacrificed the efficiency of extraction but
made it suitable for realization as a look-up table in a standard memory chip. In our implementation we use an alphabet
of $M=4$ letters and make conversions using blocks of $N=10$ such values. This matches the 20-bit address space of the
2~MB flash memory chip used. Another limitation that favors smaller alphabets and shorter processing blocks is the
requirement of the process to be stationary: if parameters of the process drift, the output sequence may be not random any more.
The process must be stationary during the time required to run through all possible combinations within a processing
block, whose number is of the order of $M^N$. Thus, a substantial increase of the alphabet size and the block length
will also require to guarantee stability not for seconds, but for hours and even days, which is almost impossible to
guarantee in practice: any feedback loop in the system, e.g. for keeping the constant
count rate, should be substantially slower than this time, rendering the system to be extremely slow and impractical.

Figure~\ref{fig_extraction} shows the simulated efficiency of the extraction algorithm for a varying size of the binary
buffer $b=N\lceil \log M \rceil$ needed to store the whole processing block. The data are calculated for a uniform
distribution of the input symbols and contain traces for different
alphabet sizes $M$, which are conveniently chosen as the powers of 2. One can see that for a
chosen pair of $M=4$ and $N=10$ the algorithm generates 1.2~bits per input symbol. It is easy to note that scaling with
the memory size is rather poor: even for a 40-bit address space, which is unrealistic and raises temporal stability
concerns, the efficiency only approaches 1.8~bits per symbol. Thus, even as we sacrificed the bit generation rate for
system simplicity and better output quality, we are still not too far from potential limits of this scheme.

\begin{figure}
\centering
\includegraphics[width=\columnwidth]{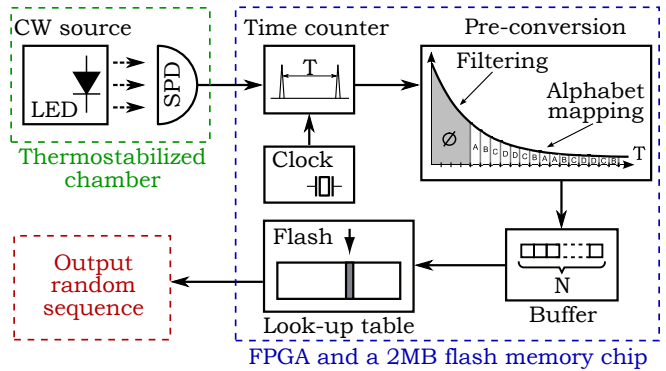}
\caption{Block diagram of the experimental setup.
}
\label{fig_setup}
\end{figure}

The discussed principles and ideas were fulfilled in our experimental realization (see Fig.~\ref{fig_setup}) based on a silicon SPD with a thin
depletion layer and a $\diameter 30$~$\mu$m sensitive area. All processing is made in an FPGA attached to a 2~MB flash
memory chip. A red LED ($\lambda \approx 627$~nm, spectral width $\Delta \lambda \approx 45$nm) driven with a $\approx 10$~$\mu$A
current is used as a light source, while the whole pair of the LED and the SPD is temperature stabilized at
+25\degree~C. A feedback loop is used for stabilization of the count rate by adjustment of the LED current. The
last one has a time constant of 16~s to ensure that the process is stationary on the time scale of $M^N$ counts.

A major flaw in the experimental system with respect to the theory is non-ideal characteristics of the SPD. For the
presented QRNG structure it is, first of all, temporal parameters of the SPD. While the photon arrival process is the
Poissonian one, the generation of SPD clicks is not. It has a highly suppressed probability of detection right after
the previous click, known as a dead time. It may also have the opposite effect with another temporal shape called
afterpulsing, i.e. triggering further detector clicks by a previous one even when no other photons present.

\begin{figure}
\centering
\includegraphics[width=\columnwidth]{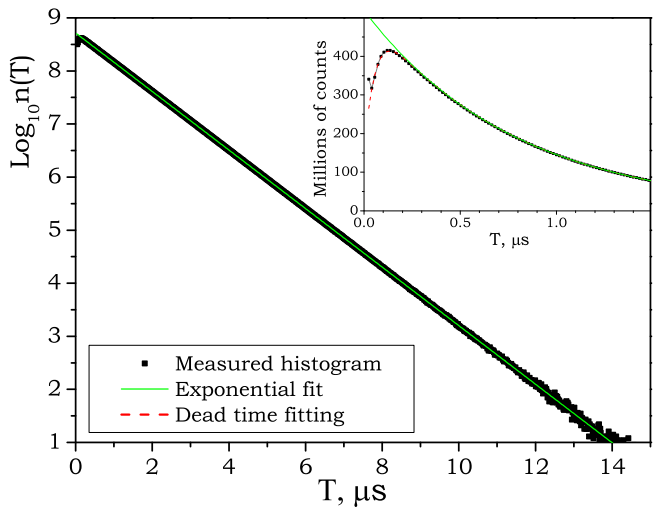}
\caption{Measured histogram of waiting times between detector clicks at the average rate of $1.2\times10^6$
clicks/s. A pronounced deviation from the expected exponential behavior is seen at $T\lesssim150$~ns.
}
\label{fig_hist}
\end{figure}

Our study of the actual process was performed at the same average count rate as used in the final device, namely
1.2~MHz. Figure~\ref{fig_hist} shows count frequency histogram as a function of a delay between successive clicks.
It ideally fits the expected exponential distribution except for the intervals shorter than 150~ns.
A deviation from the expected distribution fits well by an exponential decay function with the time constant of 40~ns.

Whatever is the nature of these non-idealities, as long as we can measure the maximal time period which is still
affected, we can easily filter out dependent events by making sure that not a single time interval shorter than the
specified time is used for random number generation. In general, we model the real SPD as an ideal detector giving
Poissonian statistics, with a perturbation function with a maximal memory of $\tau$.  To get rid of dependent events we
use a simple digital filter that rejects all time intervals shorter than 160~ns, reducing the events rate from 1.2 to
1.0~MHz.

Another significant flaw of the actual system is the presence of dark counts. The dark count rate of the detector used is
around 200~Hz, which is almost 4 orders of magnitude smaller than regular count rate. Although the nature of dark counts
is much more complicated and not obviously as ``quantum'' as photodetection events, it still has nearly Poissonian
statistics and does not affect any obtained results. We can only say that 0.01\% of the generated entropy may not be enough
``quantum''.

\begin{figure}
\centering
\includegraphics[width=0.8\columnwidth]{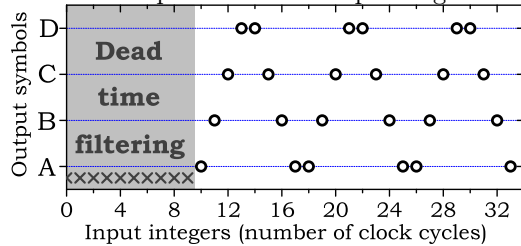}
\caption{Conversion of digitized time intervals into a 4-letter alphabet with bias reduction via bi-directional coding
and a time filter cutoff of 10 clock cycles.
}
\label{fig_mapping}
\end{figure}

The detector clicks are time tagged with a resolution of 16~ns and corresponding time intervals are calculated. All
intervals shorter than 160~ns are discarded, while the remaining ones are converted into a 4-symbol alphabet as shown in
Fig.~\ref{fig_mapping}. The particular conversion scheme gives a more uniform distribution than the trivial modulo
operation. Blocks of
10 successive symbols represented as ten 2-bit strings form a 20-bit address in the memory chip pointing to a 2-byte
pre-calculated value. The number of output bits varies depending on the symbols obtained from zero (when all 10 symbols
are the same) to 14 bits (the maximum number of permutations is $10!/(3!3!2!2!) = 25200$). To enable this variable size
output, a simple binary coding is used when the actual output data is left-padded with 1 and zeroes to form a 16-bit string
$\underbrace{\mathstrut0\dots0}_{15-k}1\underbrace{\mathstrut ab\dots yz}_k$, where $k$ is the output string size and $ab\dots yz$ is the
string itself.

Generated random bit sequences were characterized using the NIST statistical test suite. Testing 1~Gbit consecutive data
chunks with $\alpha=0.01$ using 1000 bit streams by $10^6$~bits showed the pass ratio well above 0.98 for all tests. The
P-value$_T$ of the uniformity chi-square test among P-values obtained for each stream is 0.68, which is above the 0.0001
confidence level. All obtained results suggest that generated sequences are indistinguishable from truly random ones by the
particular tests. The broad scope of the NIST suite and clear implemented QRNG operation principles confirm that the
generated random data are of high quality and may be used in critical applications.

In conclusion, we have experimentally demonstrated a quantum random number generator based on the measurement of waiting
times in the process of photon arrival at the SPD. The main concept of the device is its simplicity, robustness, and
real-time operation.
The used deterministic randomness extractor together with a straightforward raw data processing enables adaptive random bits
extraction that guarantees output quality regardless of the actual entropy of the source.

The work was supported under the Government Future Research Fund (FPI) contract.

\end{document}